# The Implementation of an ES in a Small Firm: The case of Silk Cooperation

*Completed Research Paper*

**Chandima Wickramatunga**
Swinburne University of Technology
Melbourne, Australia
cwickramatunga@swin.edu.au

**Sachithra Lokuge**
RMIT University
Melbourne, Australia
ksplokuge@gmail.com

## Abstract

*The introduction of cloud computing has provided opportunities for small businesses to implement enterprise systems (ES) in their organizations and thereby improve their business processes. While there have been many studies focusing on ES implementation among medium-large sized firms, the factors that influence the implementations of ES in such firms are different to that of small firms. This teaching case discusses an implementation of a cloud enterprise resource planning (ERP) system in a small firm in the Asian region. The case illustrates factors that enabled successful implementation of a cloud ERP system in a small firm and the lessons learnt through this successful endeavor. The case study and the teaching notes are suitable for any undergraduate or postgraduate cohort, following a course in management information systems.*

## Introduction

The implementation of enterprise systems (ES) provide increased organizational performance, efficiency and effectiveness of their business processes and access to real-time information (Seddon et al. 2010; Sedera et al. 2016). With the increasing competition, organizations despite their size, are faced with immense pressure to implement ES to gain innovation and achieve a competitive edge (Lokuge and Sedera 2014b). Earlier, due to its high resourcefulness, only large firms mainly adopted ES (Lokuge and Sedera 2017). However, with the introduction of cloud computing, even small firms have gained the opportunity to adopt ES (Walther et al. 2018).

Although existing literature highlights ES implementation in large firms, there is only a handful of research that discusses ES implementation in small firms. Further, the innate characteristics of small firms might evoke different environment and factors for a successful implementation. As such, it is important to study ES implementation in small firms as it provides (i) a better understanding of the benefits of ES for small firms, (ii) to understand factors that determine the success/failure of ES implementations in small firms, and (iii) to inform academia of how ES are implemented in small firms. The objective of this teaching case is, therefore, to demonstrate the uniqueness of cloud ES implementation in a small firm. The case description consists of comprehensive information on pre-implementation, implementation, going-live and post-implementation stages of the cloud ES implementation.

## Background of the Company

The 'E-Silk Route Ventures (Pvt) Ltd.,' trading as 'Silk Cooperation,' (henceforth referred to as SC) is an export-oriented company for organic agriculture, food, beverage, and nutraceuticals sector. The company was started in 2014 with the intention of creating an environment to uplift the livelihood of small-scale farmers. The objective of SC is to provide global customers with premium quality commodities. The Company takes care of Supply Chain from (i) Sourcing, Manufacturing/processing, (ii) Packaging, Labelling, comprising - Total Original Brand Manufacturing (OBM) / Private Labelling, (iii) Shipping / Logistics, Clearance (for some destinations) and (iv) last-mile delivery/fulfilment.

Since 2014, the company had been carrying out operations at a rented workplace and in 2019 they moved to a place of their own. The company has outsourced all their manufacturing functionalities, which enabled the company to focus on their core competency of managing the supply chains of their customers. The company has been maintaining a good supplier-relationship management with a range of suppliers across the globe.

In 2019, SC implemented a cloud-based ES to increase the efficiency and productivity of their business processes. At the time of the implementation of the system, SC had 20 employees which consisted of 5 managers and 15 executives. Before the implementation of the ES, SC had been using a simple project management software. This project management software included basic functions such as chat, document sharing and task assignment of the employees. However, with the high demand and increase in sales, the owner of the company decided that it was best that they implement an ES in the company.

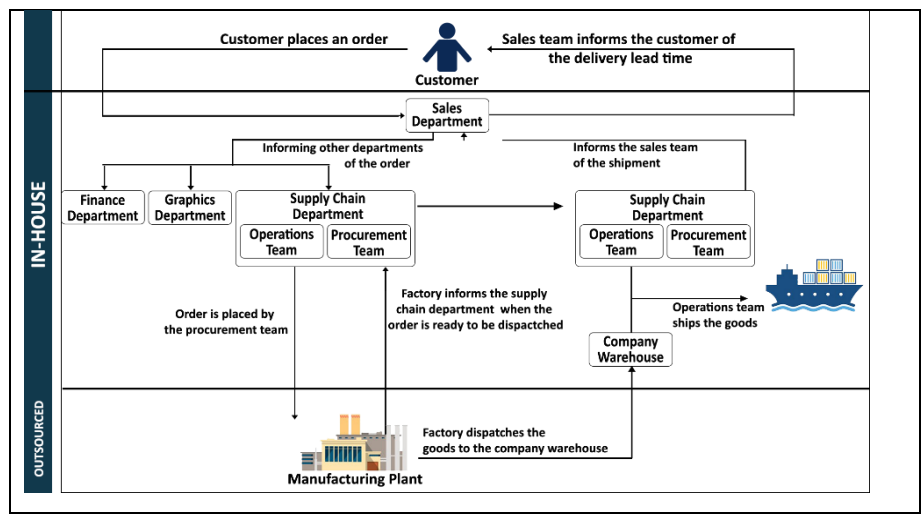

**Figure 1. Business Processes of the Company**

Figure 1 provides an overview of the business operations of SC. The company consisted of six departments such as sales and marketing, supply chain and operations, finance, IT and graphics, research and development and human resources. The ES was expected to automate the customer relationship management (CRM), the processes within the human resource department and help supply chain team handle shipments with project management tools. While the existing business processes of the company was not complex, there were many inconsistencies in the current business processes that caused delays. Each of the business function that was inconsistent required manual attention.

## Pre-implementation Stage

Since SC was mainly involved in managing supply chains, the sales team was required to provide updates on the whereabouts of the shipments to the customers. Hence, coordination between the sales and the supply chain department was crucial. Representing the total shipment/order on a Gantt chart helped the departments to properly map out the order and provide the customers with a proper timeline. Furthermore, a Gantt chart enabled the adjustment of the timeline very easily.

*"I wanted to represent the shipments visually. I thought that Gantt charts would be good since I had experience in using Gantt charts at my previous workplace"– The supply chain executive.*

The team at SC was young and consisted of tech savvy individuals. They were on the lookout for systems and tools that would help SC. They even tested a few systems by themselves to see whether they would suit SC requirement.

All customer related details pertaining to an order were entered in google sheets and this acted as the customer database which was referred by every department in the company.

*"I was entering the same customer's name multiple times. It was inefficient"– The sales executive.*

Considering theses inefficiencies, SC identified that CRM system was an essential tool for the company.

SC operation were carried out at a rented house and there was limited space. Therefore, the employees' workstations were a few feet apart. They constantly talked to the person sitting next to them to get updates. Such inefficient processes and lack of transparency of business processes therefore caused many inefficiencies at SC.

*"We were always behind the supply chain team for the updates on the shipment. I am sure that they might have found it annoying, but we were doing our job" – The sales executive.*

Information flow between departments were not streamlined and there was no transparency of transactions, since there was no proper system that integrated all the business functions together. There was no controlling and supervising mechanism for the managers as well. Files were stored in personal computers of the employees or in Google Drive. The owner was concerned regarding this matter and wanted all the information to be in one space. The Director was aware of cloud computing technology and he was keen on getting this to the company.

*"Cloud computing has enabled businesses like us to implement software much easily. This is technology that we have to make use of if we are to grow" – The owner.*

The project was aimed at overcoming the transparency and privacy related issues and to increase the productivity of the organization. The project was initiated with the direction of the company owner. During the initial discussions, it was forecasted that it would take a 3-4 months for the system implementation and another 3-6 months for the employees to understand and get used to the system.

### *Selecting the Team Members and the Timeline*

The implementation team members were selected to continue and assist with the ES implementation. While there are 20 employees in the company, only 3 were appointed to work on the ES implementation. The implementation team members were selected by the director based on their position and potential. It is important to note that none of the selected employees had prior experience in IS or IS implementations.

*"I wanted the research and development manager and supply chain executive. None of them had prior experience. But I was confident in the project team that we could make it work" – The owner.*

Time management was crucial since the director did not want to allocate any time from work hours for the implementation of the project. Although the new system was much needed, daily work and tasks were not neglected. Hence, the team mostly worked on the project after working hours.

The implementation was carried out in a systematic process. The total project took approximately 7 months to be completed. The figure below represents a timeline of the implementation process.

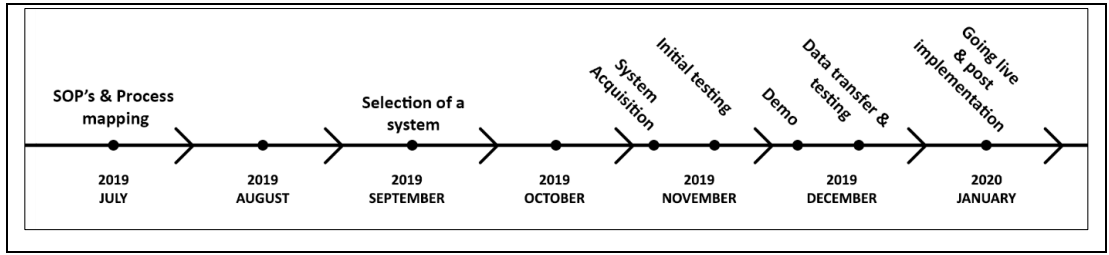

**Figure 1. ES Project Timeline**

*Creating Standard Operating Procedures (SOP) and Process Mapping*

Before the implementation, the team created SOPs for each activity and mapped the processes by using a flow chart. The ultimate objective was to identify bottlenecks in the current business process. The Human Resource Manager and the Supply Chain Executive were responsible for overlooking this process. This process was carried out in 2 stages. The initial stage, which included writing down the SOPs for each task was projected to take only 1 week.

*"We did not think that writing down SOPs would take a long time. But it took us nearly 4 weeks for everything to be completed" – The Supply Chain executive.*

The second stage included drawing of process maps with the use of an open-source software called 'Draw.io.' This process too took approximately 4 weeks to be completed. The director made sure that all the processes were in order.

*System Selection*

Once the initial step of creating and mapping out processes were completed, the project team proceed to search for a proper ES that would suit the company. The selection of a system was carried out based on the following criteria: (i) Cost, (ii) Cloud/ On Premise, (iii) Functions/ Services offered, (iv) Reviews on the web, (v) Organizational fit, (vi) Ease of use (Interface), and (vii) Vendor support. A research carried out by the research and development manager found that there were two software that would suit the company's requirements. They were 'Wrike' and 'Bitrix24.' The Table 1 below provides a comparison of Wrike and Bitrix24 based on the selection criteria.

**Table 1. An analysis between Bitrix24 and Wrike**

| Factors | Cost | Cloud based | Functionality | Reviews | Organizational fit | Difficulty | Vendor support |
|---|---|---|---|---|---|---|---|
| Bitrix24 | 4.5/5 | 5/5 | 4/5 | 4/5 | 4/5 | 4/5 | 4.2/5 |
| Wrike | 4/5 | 5/5 | 3.6/5 | 4.4/5 | 3/5 | 3/5 | 4/5 |

The project team assessed the two software and based on the above criteria, Bitrix24 was selected since it was considered less costly at the time of acquisition and had a good CRM function. However, the final decision on selecting the software was taken by the owner. System selection was conducted after 8 weeks of the first step as the company was focused on an international event that they participated; hence the whole process was delayed.

*System Acquisition*

The whole acquisition process took approximately 1 week. After the selection of a proper system, the director decided that it needs to be tested before the purchase. A date was decided for the acquisition and the director advised the project team to research about how an ES is implemented in a small firm.

The acquisition was carried out towards the end of the 2<sup>nd</sup> week of November. SC did not intend to seek external support, and therefore, their plan was to find out how to implement a system using videos that were available on the Internet. Furthermore, the vendor, i.e., Bitrix24 have tutorials uploaded on YouTube, which helped the team during the implementation.

The owner registered for a free trail before the purchase.

*"I thought that it would be best to register for a free trial first since it offered full functionality for 1 month. The other 2 agreed with me as well. I wanted to see whether this was the system that we were looking for before spending."- The owner.*

Although the team saw pictures and viewed tutorials before, the actual use of the software was difficult. It took around a day for the team to get used to the system. But the director was satisfied with what Bitrix24 offered.

*"I was satisfied with the functionality. I asked the others what they thought of the system. They were okay too. Without any hesitation, I went on to acquiring the system" – The owner.*

Seasonal discounts and offers were available for SC at the time of purchasing the software and this too was a major factor for a quick purchase decision.

*"There was a seasonal discount offered at that time. I wanted to purchase the system during that period so that we were able to make use of the discount" – The owner.*

The professional plan (software version) was purchased which did not have any restrictions on the number of users. Bitrix24 offers both cloud and on-premises solution. The purchase decision of whether the product should be cloud based or on-premises was easy since the company did not have the need nor the resources to host an on-premises software solution.

## Implementation

Once the system was acquired, on the same day, the team started configuring the system. After about 7 hours from the acquisition, SC was able to configure the basic functions of the system. They renamed the ES as SC, created a company structure, investigated the daily clock in and clock out times, explored project management tools and configured a webmail. The CRM functions and task automations were complex, and they decided to work on that once they were able to configure the basics of the system.

### *Initial Testing*

The week following the acquisition was hectic for all the managerial employees since the ES was configured as per the requirement of the company. Configuration was carried out in stages. The first stage included creating a workflow for the CRM. The next stage included configuring the HR function. The third stage included testing out webmail function followed by exploring other modules such as file management, project management, collaboration, and time management.

*"I wanted the R&D manager and the sales manager to work together on creating the workflow for the CRM since they were capable of it. I gave them guidelines on what should be done" – The owner.*

The creation of a workflow for the CRM required to automate the tasks and processes. Ultimately, it took around two days for them to configure the system. The videos on the workshops helped them in understanding the creation of the workflows. Proper knowledge regarding the business processes and the system was mandatory to create a workflow for the CRM since every possible variation was accounted for and fed into the system. Once all the workflows were created, testing was conducted using a sample customer order. All possible scenarios were tested until the CRM was fully functional. The CRM automation took around 3-5 days.

The next phase of testing included creating workflows for the Human Resource Department. The HR manager was consulted during the creation of the HR workflows. A workflow for leave requisition was created. This process involved creating the company hierarchy within the system, which included illustrating the department heads along with their respective subordinates. The previous system had not offered a functionality to apply leave through the system. Developing a company hierarchy was needed since automating leave approval required the consent of the respective department head. Upon receiving the approval from the Head, the HR manager and the director would sign off, respectively. The new process was much easy and therefore physical forms were not needed anymore and an employee could request for leave remotely.

After the automation, the webmail function was fully tested. Webmail offered the function of creating a customer deal through an email which was more time saving. The testing phase took a week to be completed.

### *Demonstration of the System*

After making sure that the CRM was running perfectly, the director decided that it was time for the system to be introduced to the rest of the employees. The director informed that a demonstration of the new system would be held within the week. Furthermore, he asked all the employees to watch some introductory videos about Bitrix24 before the demonstration so that they would understand what will be demonstrated and presented. The sales and the R&D manager were tasked with carrying out the demo since it had been them who were mainly involved in the creation of workflows.

*"I wanted every function to be running smoothly before the demonstration to the entire team" – The owner.*

The demo was carried out when all the employees were present. It went for around 1.5 – 2 hours, which was a bit longer than expected. The demo was carried out in a systematic process. The interface was thoroughly explained before going into the functions. After the introduction of the interface, common functions that everyone would use were explained followed by the explanation of the CRM. The process of creating a customer order was explained and a small demonstration of how a customer order would go through each department was shown, starting from the sales team, and then onto the Finance, Procurement and then onto Operations. Thereafter, leave requisition was explained. Once all the workflows were explained, employees were asked not to make any changes to the automations as this would disrupt the workflow.

### *Data Transfer and Final Testing*

The data transfer was carried out after the demonstration. The subscription payment for the existing system was ending, and the director did not want to spend money on two systems.

*"I did not want to spend on two systems. I was sure that the transition to the new system would not take time. I asked every employee to go through all the projects in the existing system and then transfer the important files to their computer. And the files that were not needed were grouped and archived in my computer. I advised the department heads to overlook the whole process" – The owner.*

With the departmental heads being responsible for data transfer, each employee searched the existing system for their projects, files, and the spreadsheets they were responsible for. This was a tedious task since the employees had to search every file and folder in the existing system. The projects and sheets that were found to be non-value adding, were grouped, and then archived. Sheets that could be merged were combined and made into one spreadsheet to minimize duplication of data. The transfer of data took approximately 3 days, which was earlier than expected.

After the data was transferred, the managers took steps to carry a mock test with the use of hypothetical data. Testing was conducted for (i) Client management - The use of the CRM, leads management, (ii)

Human Resource Information System (HRIS) – Employee directory, leave approval and rejection, news, and announcements, calendars, use of activity stream (The 'activity stream' acted as a news feed similar to the home page of Facebook), (iii) Project management – Tasks and subtasks, Gantt charts, Notifications, Kanban, Project planner, (iv) Time tracking (Bitrix24 offered a function to state at what time an employee started and finished work. This allows the company to see the number of hours an employee has been working in the company daily) and (v) Use of webmail. After the managers ensured that the above functions were working properly, the company moved on to the going live stage.

## Going Live

With one week before the dawn of the new year, the director announced that they would be migrating to the new system. Invitation links to all the members were sent and new accounts were made. All the credentials for each account were sent to the HR manager for safe keeping. However, the initial test did not go smoothly.

*"After everyone was logged in, I gave them sometime to look around the system. Afterwards, I ran a mock test of a customer order which included an order creation to an order fulfillment" – The owner.*

The director took control afterwards and explained each process one by one.

*"I took control and explained each step. I showed how different orders could be made. It made the team settled" – The owner.*

Once the employees settled down, the director advised the managers to follow on the questions that may arise.

*"I advised each employee to ask any question that they might come across when using the new system. I requested the managers to overlook how the new system would work" – The owner.*

During the 6-year time frame, this was the first time the company implemented a new system with many employees present in the company. The implementation process was a success without any major issues and the director considered this to be a big win. The new system impacted the departments in different ways. Some departments had a major impact where they were able to improve their productivity while the other departments did not have much of an impact.

**Table 2. Major Impacts of Bitrix24**

| Department | Impact | Employee Testimonials |
|---|---|---|
| Sales and Marketing | The sales team had the biggest impact. Having a CRM helped the department to generate deals quickly. Repetition of data was minimized. Email campaigns and integration with social networks was easy | *"Managing sales leads and tracking of customer orders through the CRM is very easy" – The sales manager.* |
| Supply Chain | Gantt charts were used to illustrate customer orders. The Kanban board were used to track shipments. Some google sheets were not used anymore since they were integrated into the system. | *It is efficient to manage all their orders through the CRM and how convenient the tracking on the Kanban board is" – Assistant manager operations.* |
| Human Resources | Leave requisition and approval was carried out through the system. Company structure was created. News and announcements were posted on the system. Employee engagement and efficiency was measured. | *"I am happy that I am able to track employee attendance and leave requisition through the system. No more forms!" - Previous HR manager.* |

The functionalities that were activated and went live were: (i) the activity stream, (ii) Webmail, (iii) Social Collaboration, (iv) Calendar – This detailed all deadlines, activities etc. This was very useful for each department to see what activities are due and when they are due, (v) Bitrix24 drive – Documents that were created, saved in this drive, (vi) Tasks and Projects, (vii) Document management – This was a version controlling system for the documents that were created in the system and (viii) Time tracking and daily to-dos.

## Post-Implementation

The implementation of the new system was a challenge for the company, however, getting accustomed to the new system was even difficult. As with any other system implementation, the company had a shakedown phase of 3-5 months where productivity was dropped since the employees found the new system to be a bit complex. Although SC expected to reduce the dependency on spreadsheets, they continued this due to lack of understanding of the new system.

The below figure illustrates the shakedown and the onwards and upwards phase after the implementation of the system (Markus and Tanis 2000). The figure shows the time taken by the company to operate at the productivity level as before. Point A – B shows the transition from the old system and Point B illustrates the date of going live with the new system. Towards the fourth week of March, a lockdown of the whole country was imposed by the government due to the COVID-19 pandemic and therefore, the employees were forced to work remotely from home. It took 5 weeks for the company to resume normal operations at the workplace. During this time, the employees had no choice, but get used to the system. This drove the employees to get accustomed thus, a productivity improvement was seen afterwards.

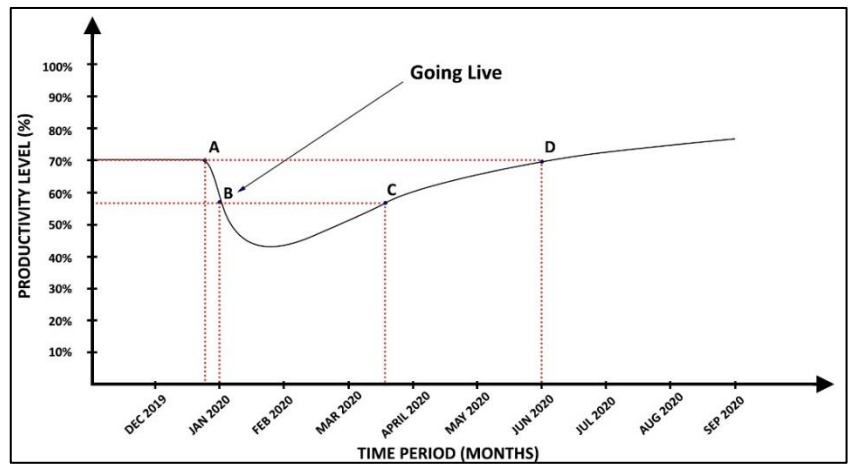

**Figure 2. Productivity Level (Pre and Post Implementation)**

Although the productivity levels were rising, the director was not satisfied since the company was not utilising the system to the maximum. The director reasoned out two main factors for the under-utilization of the system. They are as follows:

1. The company still was dependent on spreadsheets which required the manual entry of data (the employees did not seem to make an effort to switch to the system for some entry)
2. The company was at a point where they needed external support to make use of the other functions offered by Bitrix24, which required extensive knowledge on ES and coding.

### *Current Status*

It was around July 2020 when the director realized that the company was not fully utilizing the system. The company needed external support since they were at a point where they were unable to configure the rest of the system. SC did not get any support initially, due to:

1. External support being very costly, the owner did not want to spend at the point of implementation.
2. The director wanted this to be a learning process for the team as well, hence the reason for watching videos and self-learning.
3. The owner did not want another team during the implementation process which would have been intimidating for some team members.

Apparently, there were many functionalities in Bitrix24 that was not useful for SC in their day-to-day work. In addition, SC already had a customized external Inventory Management System (IMS) that ran parallel with the ES. Moreover, SC did not use in-built tools such as the Social Media Analytics. Instead, they use third party software applications. However, once the team was familiar with the system, SC decided to obtain support to further extend the system.

The reason as to why they could not execute some functions themselves was because the implementation required further technical knowledge to configure the system. Therefore, as with any other sourcing operation, the director started to request for quotations from IT companies to see whether it was feasible in getting external support. It was found that external support would be very expensive. Therefore, the director decided to recruit an experienced technical staff with knowledge on configuring cloud computing. The director expected that with the extension of their ES solution, the ES matters could be handled in-house.

*"I don't think we can use 100% of Bitrix24, but maybe 70-80% with some external systems as well; would be where we would opt for. It has too much functionality. We might not need it" – The owner.*

Currently, SC is working on their SC smoothly. The new IT executive is managing all ES related matters. Further, he is in the process of extending the functionalities of their ES.

## Conclusions: Lessons Learnt

As per Ghobadian and Gallear (1997) small firms have flexible organizational hierarchy, high top management visibility, close top management intervention, and low or no degree of formalization. Such characteristics were evident in SC as well. The case of SC demonstrates that even small firms can implement an ES quite easily in comparison to implementations conducted decades ago. This is mainly due to the advancement of cloud computing. It took about 7 months for the ES implementation process; from pre – implementation to going live. It is important to note that the implementation was done by employees without any prior ES implementation experience and with basic technical understanding. Such instances epitome the innate characteristics of cloud computing such as ease-of-use, ease-of-implementation and ease-of-learning (Lokuge et al. 2019; Sedera and Lokuge 2017; Walther et al. 2013). It was evident through the case, that top management support was the most critical success factor in the implementation process. This finding is consistent with prior literature (e.g., Lokuge 2015; Lokuge et al. 2018; Sedera 2006) on highlighting the role of the senior manager/executive in ES implementations. While ES implementation processes are structured in medium-large firms (Sedera et al. 2003a; Sedera et al. 2003b), it is also important to note that the implementation process is quite different (i.e., chaotic and complex) in small firms. Although SC had not requested external consultant support for the implementation process, it was clear that, even with new and improved technology, they were not able to fully utilize the system with the existing resources. Due to high cost of obtaining external consultancy, SC hired an IT executive for the job. The cost of the implementation was the main deciding factor when acquiring the system. The availability of cloud ES was a driving factor for the owner to consider an ES implementation. A young, knowledgeable workforce, and availability of infrastructure were some of the other factors that enabled a successful implementation.

# Teaching Notes

The teaching notes are divided into six modules. These modules can be used individually or in any combination appropriate for a course. In the six modules, we developed eleven discussions that an academic could use to discuss about ES implementation in a small firm in a developing country. In each of the eleven discussion items, we provide specific 'lesson suggestions' to demonstrate how the cases can be integrated into the curricula. Moreover, where applicable, the intersections of the modules are also provided to facilitate an integrated learning approach.

*Teaching Objectives*

1. To make students aware of the importance of an ES for a small firm in a developing country.
2. To explain how an ES implementation would take place in a small firm in comparison to a large firm.
3. To identify critical success factors ES implementation in a small firm and the role of cloud computing.

## Module 1: Overview of ES and cloud technologies

*1. What is an ES?*

ES are integrated software packages (Davenport 1998; Sedera and Lokuge 2019; Zhong Liu and Seddon 2009) that offers seamless integration of information among business functions (Davenport 1998) while automating core business processes (Holland et al. 1999; Lokuge and Sedera 2018). In the present context, firms that adopt ES, have a wide range of options for the implementation and post implementation such as; implementing the system by oneself, selective assistance or to total outsourcing (Markus and Tanis 2000).

*LESSON SUGGESTIONS: Based on the narrative above, the academics can discuss about ES benefits, whether it is a hinderance or an enabler for innovations. Further, the academics can discuss differences between cloud vs on-site ES, the differences in ES in small vs large business context.*

*2. What are cloud computing technologies?*

Cloud computing solutions have helped companies to reduce costs, improve efficiencies and enhance business processes (Lokuge and Sedera 2020; Salim et al. 2015; Walther et al. 2013). Such advantages have driven companies to focus more on utilizing cloud IT infrastructure in their organizations. This is especially applicable for small firms as the cost efficiencies and the process improvements are main factors for adopting such systems.

*LESSON SUGGESTIONS: The academics could (i) highlight the characteristics of cloud computing, (ii) benefits and limitations of cloud computing, and (iii) develop an understanding on how such technologies can be utilized in small firms.*

*3. Is the implementation of an ES the same across all types of organizations?*

As per the discussion above, does size matter for ES implementation? Students could identify how implementation of medium-large firms happen, what factors are important, whether there is a difference in public vs private firms in ES implementations (Sedera 2011; Sedera 2016; Sedera et al. 2002) etc.

*LESSON SUGGESTIONS: The academics could (i) highlight the characteristics of small vs large firms, (ii) compare large vs small firms, and (iii) develop an understanding of differences in ES implementation process with small vs large by comparing the case study findings with prior literature on large firms' ES implementations.*

## Module 2: Critical Success Factors of an ES Implementation in a Small Firm

There are many studies that have been conducted on the topic of CSF at different levels (Esteves and Pastor 2005; Nah et al. 2001; Nah et al. 2003). But there is very little literature on the CSF of an ES implementation in a small firm. It is important to identify and understand what CSF might contribute towards the successful implementation of an ES in different organizations.

*LESSON SUGGESTIONS: The academics could ask students to (i) identify CSF of ES implementation, (ii) identify CSF based on the teaching case, and (iii) identify their applicability to small firm's ES implementation.*

## Module 3: Pre – Implementation

### 4. *How has SC managed the complexity of the ES implementation?*

Mainly, it has been the direction and the strategy of the owner which has led the company to have a smooth implementation. Such involvement of the owner helped reducing the complexity in the adoption process.

*LESSON SUGGESTIONS: This is an interesting factor that the academics could discuss with students. The academic could ask the students to (i) identify problems they faced before the ES implementation, (ii) identify benefits of ES implementation for a small firm, and (iii) identify factor that small firms need to consider in order to go ahead with an ES implementation.*

## Module 4: Implementation

### 5. *Does a project team for an ES implementation require prior knowledge on implementation?*

Technical knowledge and expertise of the implementation team is considered as one of the main factors that determine success of implementation (Lech 2016; Sedera and Dey 2013). It is evident that prior training or knowledge on ES implementation is not required. However, through social media platforms the team was able to gain knowledge.

*LESSON SUGGESTIONS: The students can (i) identify issues with implementation process of the case study, and (ii) identify how they could minimize risks of such approach.*

## Module 5: Post-Implementation

### 6. *What are the factors that impacted productivity to drop after the implementation?*

As per Ross and Vitale (2000) after the ES implementation, the productivity drop is inevitable. The concept of ES was new to some employees and more learning and development were required since ES are very complex systems (Lokuge and Sedera 2014a; Lokuge and Sedera 2014c). Hence, it is common for most organizations to have a productivity drop after the implementation.

*LESSON SUGGESTIONS: The students can (i) identify ways to minimize this drop in productivity, and (ii) identify how they could guarantee a smooth transition of business processes.*

### 7. *Is External support mandatory when implementing an ES?*

From the case of SC, it can be understood that external support will be a requirement post-implementation since there might be issues that may arise while using the new system.

*LESSON SUGGESTIONS: The academic could invigorate a discussion on (i) the role of the consultants (Nuwangi et al. 2012), (ii) advantages and disadvantages of utilizing consultants for ES implementation, and (iii) what could SC do better in terms of getting support for the ES implementation.*

8. *Using the sample case, develop a generic framework for the ES implementation process at SC.*

Considering the ES lifecycle, determine a framework for ES implementations in small firms.

*LESSON SUGGESTIONS: The academic could ask the students to (i) determine high level activities of ES implementation, and (ii) identify the role of CSF in each of these activities.*

## Module 6: Theoretical Discussion

Theories would provide generalizable and broad application guidelines for understanding ES implementation in small firms. It also attempts to explain or predicts events or situations by specifying relations among variables (Dubin 1978).

9. *Dynamic capabilities and Resource Based View.*

Originally RBV was proposed by Penrose (1959), where he posits that a firm is a bundle of resources that require continuous management. Barney (1991) describe that "resources include assets, capabilities, processes, attributes, knowledge and know-how that are possessed by a firm and that can be used to formulate and implement competitive strategies." As such, technologies can be viewed as a resource that firms possess. As Grant (1991, p. 122) states "the types, the amounts of the resources available to the firm have an important bearing on what the firm can do." The Dynamic Capabilities (DC) View is an extension of the RBV Theory and has sparked great interest over the years which highlights the need to change and adapt in the face of changing business environments (Teece et al. 1997). Helfat and Winter (2011) states that a DC is explained as "one that enables a firm to alter how it currently makes its living." The academics can use these theories to understand the benefits of implementing ES in a small firm and see how such assets and capabilities differ in a small firm.

10. *IS Success model and Knowledge Based Theory*

The DeLone and McLean IS success model is one of the popular and most validated IS success models (Delone and McLean 2003). The IS success model was extended by Gable et al. (2003). The theory aims to provide a deep understanding of IS success. The Knowledge based view states that the services given by tangible resources will depend on how they are combined and applied, thereby functioning as the firm's knowledge (Alavi and Leidner 2001). Since knowledge is difficult to imitate, the knowledge based view states that these assets can produce long term sustainable competitive advantage (Sedera and Gable 2010). The application of these two theories can guide students to understand IS success factor applicable for a small firm and how knowledge management factors differ in a small firm.

11. *Technology Acceptance model*

The technology acceptance model (TAM), proposed by Davis, can be utilized to understand the adoption behavior of the small firms (Chuttur 2009; Davis 1989). Academics can guide students to understand the factors that determine the ES adoption in small firms.

## References

Alavi, M., and Leidner, D. 2001. "Review: Knowledge Management and Knowledge Management Systems: Conceptual Foundations and Research Issues," *MIS quarterly* (1), 03/01, p 107.

Barney, J. 1991. "Firm Resources and Sustained Competitive Advantage," *Journal of Management* (17:1), pp 99-120.

Chuttur, M. 2009. "Overview of the Technology Acceptance Model: Origins, Developments and Future Directions," *All Sprouts Content* (9), 01/01.

Davenport, T. 1998. "Putting the Enterprise into the Enterprise System," in: *Harvard business review*. pp. 121 - 131.

Davis, F.D. 1989. "Perceived Usefulness, Perceived Ease of Use, and User Acceptance of Information Technology," *MIS Quarterly* (13:3), pp 319-340.

Delone, W.H., and McLean, E.R. 2003. "The Delone and Mclean Model of Information Systems Success: A Ten-Year Update," *Journal of Management Information Systems* (19:4), pp 9-30.

Dubin, R. 1978. *Theory Building*. New York, N.Y: The Free Press.

Esteves, J., and Pastor, J. 2005. "A Critical Success Factor's Relevance Model for Sap Implementation Projects," in: *Managing Business with Sap,* M. Khosrow-Pour (ed.). Hershey, PA: Idea Group Publishing, pp. 240-262.

Gable, G., Sedera, D., and Chan, T. 2003. "Enterprise Systems Success: A Measurement Model," *Proceedings of the 24$^{th}$ International Conference on Information Systems* S.T. March, A. Massey and J.I. DeGross (eds.), Seattle, Washington: Association for Information Systems, pp. 576-591.

Ghobadian, A., and Gallear, D. 1997. "Tqm and Organization Size," *International Journal of Operations & Production Management* (17:2), pp 121-163.

Grant, R. 1991. "The Resource-Based Theory of Competitive Advantage: Implications for Strategy Formulation," *California Management Review* (33:3), pp 114-135.

Helfat, C.E., and Winter, S.G. 2011. "Untangling Dynamic and Operational Capabilities: Strategy for the (N)Ever-Changing World," *Strategic Management Journal* (32:11), pp 1243-1250.

Holland, C., Light, B., and Gibson, N. 1999. *A Critical Success Factors Model for Enterprise Resource Planning Implementation*.

Lech, P. 2016. "Causes and Remedies for the Dominant Risk Factors in Enterprise System Implementation Projects: The Consultants' Perspective," *SpringerPlus* (5:1), February 29, p 238.

Lokuge, K.S.P. 2015. "Agile Innovation: Innovating with Enterprise Systems," in: *Information Systems School*. QUT ePrints: Queensland University of Technology.

Lokuge, S., and Sedera, D. 2014a. "Deriving Information Systems Innovation Execution Mechanisms," *Australasian Conference on Information Systems*, Auckland, New Zealand: AIS.

Lokuge, S., and Sedera, D. 2014b. "Enterprise Systems Lifecycle-Wide Innovation," *Americas Conference on Information Systems (AMCIS 2014)*, Savannah, Georgia: AIS.

Lokuge, S., and Sedera, D. 2014c. "Enterprise Systems Lifecycle-Wide Innovation Readiness," *Pacific Asia Conference on Information Systems*, Chengdu, China: AIS.

Lokuge, S., and Sedera, D. 2017. "Turning Dust to Gold: How to Increase Inimitability of Enterprise System," *Pacific Asia Conference on Information Systems*, Langkawi, Malaysia: AIS.

Lokuge, S., and Sedera, D. 2018. "The Role of Enterprise Systems in Fostering Innovation in Contemporary Firms," *Journal of Information Technology Theory and Application (JITTA)* (19:2), pp 7-30.

Lokuge, S., and Sedera, D. 2020. "Fifty Shades of Digital Innovation: How Firms Innovate with Digital Technologies," *Pacific Asia Conference on Information Systems*, Dubai, UAE: AIS, p. 91.

Lokuge, S., Sedera, D., Grover, V., and Xu, D. 2019. "Organizational Readiness for Digital Innovation: Development and Empirical Calibration of a Construct," *Information & Management* (56:3), pp 445-461.

Lokuge, S., Sedera, D., and Perera, M. 2018. "The Clash of the Leaders: The Intermix of Leadership Styles for Resource Bundling," *Pacific Asia Conference on Information Systems*, Yokohama, Japan: AIS.

Markus, L., and Tanis, C. 2000. "The Enterprise Systems Experience - from Adoption to Success," in: *Framing the Domains of It Management: Projecting the Future through the Past,* R.W. Zmud (ed.). Cincinnati, OH: Pinnaflex Educational Resources, Inc, pp. 173-207.

Nah, F.F.-H., Lau, J.L.-S., and Kuang, J. 2001. "Critical Factors for Successful Implementation of Enterprise Systems," *Business Process Management Journal* (7:3), pp 285-296.

Nah, F.F., Zuckweiler, K.M., and Lau, J.L. 2003. "Erp Implementation: Chief Information Officers' Perceptions of Critical Success Factors," *International Journal of Human-Computer Interaction* (16:1), pp 5-22.

Nuwangi, S.M., Sedera, D., and Murphy, G. 2012. "Multi-Level Knowledge Transfer in Software Development Outsourcing Projects: The Agency Theory View," in: *International Conference on Information Systems*. Orlando, Florida.

Penrose, E.T. 1959. *The Theory of the Growth Ofthe Firm*. New York: Sharpe.

Ross, J.W., and Vitale, M.R. 2000. "The Erp Revolution: Surviving Vs. Thriving," *Information Systems Frontiers* (2:2), pp 233-241.

Salim, S.A., Sedera, D., Sawang, S., Alarifi, A.H.E., and Atapattu, M. 2015. "Moving from Evaluation to Trial: How Do Smes Start Adopting Cloud Erp?," *Australasian Journal of Information Systems* (19), pp S219-S254.

Seddon, P.B., Calvert, C., and Yang, S. 2010. "A Multi-Project Model of Key Factors Affecting Organizational Benefits from Enterprise Systems," *MIS Quarterly* (34:2), pp 305-328.

Sedera, D. 2006. "An Empirical Investigation of the Salient Characteristics of Is-Success Models," *Americas Conference on Information Systems*, Acapulco, Mexico: AIS.

Sedera, D. 2011. "Size Matters! Enterprise System Success in Medium and Large Organizations," in: *Enterprise Information Systems: Concepts, Methodologies, Tools and Applications,* I.R.M. Association (ed.). Pennsylvania, United States: IGI Global, pp. 958-971.

Sedera, D. 2016. "Does Size Matter? The Implications of Firm Size on Enterprise Systems Success," *Australasian Journal of Information Systems* (20), pp 1-25.

Sedera, D., and Dey, S. 2013. "User Expertise in Contemporary Information Systems: Conceptualization, Measurement and Application," *Information & Management* (50:8), pp 621–637

Sedera, D., Gable, G., and Chan, T. 2003a. "Erp Success: Does Organization Size Matter?," *Pacific Asia Conference on Information Systems*, Adelaide, Australia: Association for Information Systems, pp. 1075-1088.

Sedera, D., Gable, G., and Chan, T. 2003b. "Survey Design: Insights from a Public Sector-Erp Impact Study," *Pacific Asia Conference on Information Systems*, Adelaide, Australia: AIS, pp. 595-610.

Sedera, D., Gable, G., and Palmer, A. 2002. "Enterprise Resources Planning Systems Impacts: A Delphi Study of Australian Public Sector Organisations," *Pacific Asia Conference on Information Systems*, Tokyo, Japan: AIS.

Sedera, D., and Gable, G.G. 2010. "Knowledge Management Competence for Enterprise System Success," *The Journal of Strategic Information Systems* (19:4), pp 296-306.

Sedera, D., and Lokuge, S. 2017. "The Role of Enterprise Systems in Innovation in the Contemporary Organization," in: *The Routledge Companion to Management Information Systems,* R.G. Galliers and M.-K. Stein (eds.). Abingdon, United Kingdom: The Routledge p. 608.

Sedera, D., and Lokuge, S. 2019. "Does It Get Better over Time? A Longitudinal Assessment of Enterprise System User Performance," *Information Technology & People* (33:4), pp 1098-1123.

Sedera, D., Lokuge, S., Grover, V., Sarker, S., and Sarker, S. 2016. "Innovating with Enterprise Systems and Digital Platforms: A Contingent Resource-Based Theory View," *Information & Management* (53:3), pp 366–379.

Teece, D.J., Pisano, G., and Shuen, A. 1997. "Dynamic Capabilities and Strategic Management," *Strategic Management Journal* (18:7), pp 509-533.

Walther, S., Sedera, D., Sarker, S., and Eymann, T. 2013. "Evaluating Operational Cloud Enterprise System Success: An Organizational Perspective," *European Conference on Information Systems (ECIS 2013)*, Utrecht, p. 16.

Walther, S., Sedera, D., Urbach, N., Eymann, T., Otto, B., and Sarker, S. 2018. "Should We Stay, or Should We Go? Analyzing Continuance of Cloud Enterprise Systems," *Journal of Information Technology Theory and Application (JITTA)* (19:2), pp 57-88.

Zhong Liu, A., and Seddon, P.B. 2009. "Understanding How Project Critical Success Factors Affect Organizational Benefits from Enterprise Systems," *Business Process Management Journal* (15:5), pp 716-743.